\documentclass{article}
\usepackage{amsmath,graphicx,mlspconf,amssymb}
\usepackage{xcolor}
\usepackage{bbm}
\usepackage{booktabs}
\usepackage{multirow}
\usepackage{tabularx}
\usepackage{booktabs, multirow, makecell}
\usepackage{graphicx}

\usepackage[acronym]{glossaries}
\newacronym{ai}{AI}{Artificial Intelligence}
\newacronym{cnn}{CNN}{Convolutional Neural Network}
\newacronym{vad}{VAD}{Voice Activity Detection}
\newacronym{sgvad}{SG-VAD}{Stochastic Gates Based VAD}
\newacronym{rms}{RMS}{Root Mean Square}
\newacronym{gscv2}{GSCV2}{Speech Google Commands
V2 dataset}
\newacronym{roc}{ROC}{Receiver Operating Characteristic}
\newacronym{tpr}{TPR}{true positive rate}
\newacronym{fpr}{FPR}{false positive rate}
\newacronym{auc}{AUC}{Area Under the Curve}
\newacronym{snr}{SNR}{Signal-to-Noise Ratio}

\toappear{2025 IEEE International Workshop on Machine Learning for Signal Processing, Aug.\ 31-- Sep.\ 3, 2025, Istanbul, Turkey}

\title{Tiny Noise-Robust Voice Activity Detector for Voice Assistants}
\name{\parbox{\textwidth}{\centering
   Hamed Jafarzadeh Asl$^{\ \star\dagger}$
   \quad Mahsa Ghazvini Nejad$^{\ \star \dagger}$
   \quad Amin Edraki$^{\ \ddagger}$
   \\ Masoud Asgharian$^{\ \|}$
   \quad Vahid Partovi Nia$^{\ \dagger \S}$\thanks{$\star$ Equal contribution.}
       \thanks{© 2025 IEEE. Personal use of this material is permitted. Permission from IEEE must be obtained for all other uses, in any current or future media, including reprinting/republishing this material for advertising or promotional purposes, creating new collective works, for resale or redistribution to servers or lists, or reuse of any copyrighted component of this work in other works.}
}}
\address{%
   $^{\dagger}$ Huawei Noah’s Ark Lab, Montreal Research Center, Montreal, Canada \\%
   $^{\ddagger}$ Huawei Noah’s Ark Lab, Toronto Research Center, Toronto, Canada \\%
   $^{\|}$ McGill University, Montreal, Canada \qquad
   $^{\S}$ Polytechnique Montreal, Montreal, Canada 
}

\begin{document}

\maketitle

\begin{abstract}
%
\gls{vad} in the presence of background noise remains a challenging problem in speech processing. Accurate \gls{vad} is essential in automatic speech recognition, voice-to-text, conversational agents, etc, where noise can severely degrade the performance. A modern application includes the voice assistant, specially mounted on Artificial Intelligence of Things
(AIoT) devices such as cell phones, smart glasses, earbuds, etc, in which the voice signal includes background noise.  Therefore, \gls{vad} modules must remain light-weight due to their practical on-device limitation.  The existing models often struggle with low signal-to-noise ratios across diverse acoustic environments. A simple \gls{vad} often detects human voice in a clean environment, but struggles to detect the human voice in noisy conditions. We propose a noise-robust \gls{vad} that comprises a light-weight \gls{vad}, with data pre-processing and post-processing modules to handle the background noise. This approach significantly enhances the \gls{vad} accuracy in noisy environments and requires neither a larger model, nor fine-tuning. Experimental results demonstrate that our approach achieves a notable improvement compared to baselines, particularly in environments with high background noise interference. This modified \gls{vad}  additionally improves clean speech detection. 
\end{abstract}

\section{Introduction}
\label{sec:intro}

Voice activity detection (VAD) is a crucial task in speech processing, aimed at distinguishing between speech and non-speech segments in an audio signal. \gls{vad} plays a vital role in enhancing the performance of speech recognition systems, noise reduction, speech coding, and other speech-based applications. With the growing use of portable Artificial Intelligence of Things (AIoT) devices, \gls{vad} has become increasingly important, as these devices often operate in challenging acoustic conditions, such as crowded public spaces or moving vehicles. Accurately detecting speech in such environments while maintaining low computational cost and low power consumption remains a significant challenge for real-time applications. 

\begin{figure}[t!]
    \centering
    \includegraphics[width=.8\linewidth]{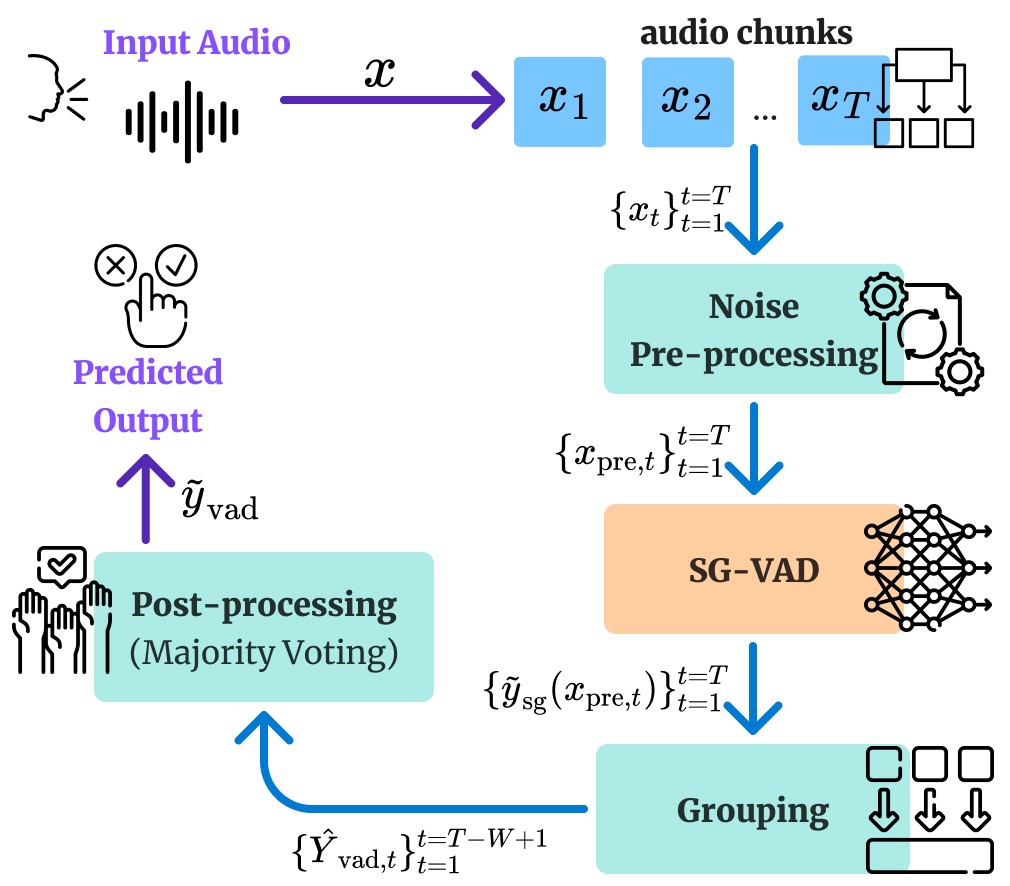}
    \caption{Block diagram of the proposed \gls{vad} pipeline, including segmentation, noise-removal pre-processing, inference with \gls{sgvad}, sliding-window grouping, and post-processing using majority voting, see Section~\ref{sec:vad_pipeline}.}
    \label{fig:vad_pipeline}
\end{figure}

\gls{vad} has evolved significantly over the past years, beginning with classical signal processing techniques. Early methods relied on simple features like energy, zero-crossing rate, and spectral entropy to detect speech presence \cite{benyassine1997itu}. These approaches, though effective in controlled environments, struggle in noisy conditions. More recent advancements have integrated deep learning with signal processing, gaining prominence in the last few years. Such methods significantly improved \gls{vad} accuracy, particularly in complex, real-world noise environments \cite{sehgal2018convolutional}.

Many state-of-the-art \gls{vad} models, especially deep learning-based approaches, are too large for deployment on edge devices. This limitation raises interest in lightweight \gls{vad} systems that are efficient under power and memory constraints. However, many lightweight models are designed for clean speech versus silence detection, making them less effective in noisy speech mixtures.
This work focuses on improving \gls{vad} performance in such challenging noisy environments.

Voice assistant is a key feature in common edge devices. \gls{vad} serves as one of the first steps in processing audio inputs. \gls{vad} identifies speech segments from background noise, ensuring that only speech data is sent for further processing. This step is crucial for reducing computational load and enhancing the accuracy of subsequent stages, such as speech recognition and natural language understanding. 

Recent research has seen the development of several lightweight DNN-based \gls{vad} architectures aimed at improving both efficiency and accuracy. For instance, dilated convolutions have been used in \cite{chang2018temporal} to capture long-term temporal dependencies, without increasing model complexity. Alternatively, in \cite{braun2021training}, gated recurrent units were employed to model temporal dependencies. These architectures emphasize the trade-off between capturing complex features and maintaining a low computational cost.


A notable contribution to lightweight \gls{vad} design is MarbleNet \cite{jia2021marblenet}, which employs depthwise separable convolutions to reduce computational overhead, while maintaining performance. \gls{sgvad} \cite{SG_VAD_main_paper} is another notable example of lightweight \gls{vad} architectures and is based on one-dimensional time-channel separable convolutions and the locally stochastic gates mechanism. \gls{sgvad} is  introduced in detail in Section~\ref{sec:background} and used as our primary baseline. Our proposed methodology is explained in Sections~\ref{sec:postprocessing}--\ref{sec:vad_pipeline}.




\section{Background}
\label{sec:background}


Having formulated voice activity detection as a denoising task, Svirsky and Lindenbaum~\cite{SG_VAD_main_paper} introduced the \gls{sgvad} model that aims to identify noise patterns in audio segments. \gls{sgvad} divides the input waveform $x$ into $D$ non-overlapping time frames denoted by $x\!=\!\left\{ x_{d} \right\}_{d=1}^{D}$, and predicts the \gls{vad} label (speech vs. non-speech), $\tilde{y}_{\text{sg}}(x)$, as follows
\vspace{-0.8em}
\begin{equation}
\label{sgvad_inference_procedure}
\begin{split}
\hat{y}_{\text{sg}}(x) & = \frac{1}{D}\sum_{d=1}^{D}{\sum_{c=1}^{C}{z_{d,c}}}, \\
    \tilde{y}_{\text{sg}}(x) & = \mathbbm{1}
_{\{\hat{y}_{\text{sg}}(x) \geq \mathrm{thresh}\}}~,
\end{split}
\end{equation}
where $D$ is total number of non-overlapping time frames, $C$ is the number of frequency channels, $z_{d,c}$ denotes the output of \gls{sgvad} for $d$-th time frame and $c$-th frequency channel, and $\mathbbm{1}_{\{\hat{y}_{\text{sg}}(x) \geq \mathrm{thresh}\}}$ is the indicator function on the set $\{\hat{y}_{\text{sg}}(x) \geq\mathrm{thresh}\}$ to produce \gls{vad} labels.
Although \gls{sgvad} performs well in detecting clean speech versus non-speech,  it performs poorly in real-world scenarios. Voice assistants run on portable devices are commonly used in noisy environments where noise-robust \gls{vad} is necessary.



\gls{sgvad} is trained on a dataset that contains separate speech and noise samples. Speech samples in the dataset are extracted from \gls{gscv2} ~\cite{gscv2_dataset}, and noise segments are obtained from FS2K ~\cite{fs2k_dataset}. \gls{sgvad} detects frequency-domain noise patterns. Therefore it requires a dataset consisting of separate speech and noise samples. However, their model must be evaluated on real-world scenarios of speech and noise combined datasets, eventually for practical reasons.

\gls{sgvad} performance drops considerably when tested on datasets containing noisy speech samples, which will be discussed in more details in Section~\ref{sec:results}. 
A straightforward solution is to fine-tune the existing \gls{sgvad} model using a dataset that includes noisy speech instances. This solution faces two important challenges. First, fine-tuning requires accurate and high-resolution labelled data. For training \gls{sgvad}, small segments of audio signals (less than a second long) should be labelled as speech or noise, which is challenging. 
The second challenge of fine-tuning \gls{sgvad} comes from the model size. Since \gls{sgvad} is a small model with only 7.8K parameters designed to detect noise-only versus clean speech patterns, it may not be expressive enough to differentiate noisy speech from noise-only segments, particularly in low \gls{snr} conditions. Therefore, fine-tuning \gls{sgvad} for noisy environments may require an increased number of model parameters, compromising one of the key advantages of \gls{sgvad}: its small model size.

With the above-mentioned considerations in mind, we propose to augment the pre-trained inference framework of \gls{sgvad} with pre-processing and post-processing modules, enhancing the model's ability to identify speech in noisy conditions, without re-training or fine-tuning the model. While only evaluated using the \gls{sgvad} model, the proposed augmentations can be used to modify similar \gls{vad} models to boost their performance in the presence of background noise.

\section{Post-processing}
\label{sec:postprocessing}


\gls{sgvad} performs well in detecting clean and continuous speech during inference. Equation \eqref{sgvad_inference_procedure} demonstrates that \gls{sgvad} predicts the \gls{vad} label for an audio input by averaging the per-frame and per-frequency \gls{vad} scores across the entire input waveform and all frequency bins. In other words, speech is detected when the speech signal is present in most time frames. Although this method shows reasonable performance on the datasets tested in ~\cite{SG_VAD_main_paper}, it lacks a real-world evaluation scenario in which the speech signals include background noise. The AVA speech~\cite{ava_speech_dataset} and HAVIC~\cite{havic_dataset} datasets, which are used to evaluate the \gls{sgvad} model, consist of separate noise and speech instances. In the speech-labelled samples, speech is present continuously in the majority of the time frames. However, voice assistants often operate while the user does not speak continuously, such as issuing short commands or pausing between phrases~\cite{non-continuous_speaking_supporting_reference}.

We propose an augmented \gls{vad} with post-processing method that relaxes the continuity of the speech signal over the duration of the input sample. We divide the input signal into fixed-length segments and find the \gls{sgvad} decision 
$\tilde{y}_{\text{sg}}(x_t)$  for each time frame $1\leq t \leq T$. Then, a moving window of length $W\!<\!T$ is built so that the \gls{vad} label of the input is predicted based on this moving window
\begin{equation}
    \hat{Y}_{\text{vad},t}=\left\{ \tilde{y}_{\text{sg}}(x_{k}) \right\}_{k=t}^{k=t+W-1},
\end{equation}
where $1\leq t \leq T\!-\!W\!+\!1$ indexes the moving window. The proposed speech detection criterion simply takes a majority vote over each window. In another word \gls{vad} detects speech in each window when majority of \gls{vad} decisions recognize speech-related patterns in the input audio. The overall \gls{vad} decision is based on detecting at least a single window with speech signal presence. If speech is detected in any of the windows, the algorithm signals speech detection in the sequence. This approach resolves the problem of non-continuous speech and short commands, see Section~\ref{sec:vad_pipeline} for more details.

\section{Pre-processing}

\label{sec:preprocessing}

The post-processing method explained in Section~\ref{sec:postprocessing} only tackles the continuity issue, not the \gls{vad} problem in presence of noise. We propose to employ three pre-processing methods applied to the input audio signal, aiming to improve \gls{vad} performance in noisy environments.

\subsection{Spectral Subtraction}
\label{sec:spectral_subtraction}

Spectral subtraction is a widely used frequency-domain method for noise reduction, especially in speech processing~\cite{spectral_subtraction_reference_1,spectral_subtraction_reference_2}. The idea behind this approach is to estimate the power spectrum of the background noise and subtract it from the noisy speech spectrum. Spectral subtraction performs best in reducing additive stationary noise.

Given a noisy audio signal in the frequency domain, $X(f)$, and an estimated noise spectrum, $\hat{N}(f)$, the magnitude of the clean signal spectrum $|\hat{S}(f)|$ is estimated~as
\begin{equation}
\label{formula:spectral_subtraction}
|\hat{S}(f)| = \max\left({|X(f)| - \alpha|\hat{N}(f)|},~\beta|\hat{N}(f)| \right),
\end{equation}
where $\alpha$ is the over-subtraction factor and $\beta$ is the spectral floor parameter. The over-subtraction factor  $\alpha\!>\!1$ compensates for noise underestimation, while $\beta\!\in\![0,\!1]$ avoids negative or very small values that cause undesired signal distortions~\cite{spectral_subtraction_reference_3}. An estimate of the clean signal is reconstructed using the noisy speech signal phase.

\subsection{Energy Gating}
\label{sec:energy_gating}

Energy gating, also known as noise gating, is a time-domain, frame-based signal enhancement technique that attempts to suppress or attenuate portions of the signal whose energy falls below a predefined threshold. By computing the short-time energy of overlapping frames, the method selectively retains segments with sufficient energy, while pointing or discarding low-energy frames, typically considered to be background noise or silence~\cite{energy_gating_reference_1,energy_gating_reference_2}.

Let $x[n]$ denote the input time-domain signal, $x_m[n]$ be the $m$-th frame with $N$ samples, and $E_m$ define the energy of $x_m[n]$ as $E_m\!=\!\sum_{n=1}^{N}\! x_m^2[n]$. Hence, the energy gating function can be formulated as
\begin{equation}
\label{formula:energy_gating}
\tilde{x}_m[n] = 
\begin{cases}
x_m[n], & \text{if } E_m \geq \theta, \\
0, & \text{if } E_m < \theta,
\end{cases}
\end{equation}
where $\theta$ is the energy threshold value. Here, $\tilde{x}[n]\!=\!\sum_m \tilde{x}_m[n]$ represents the output signal after applying the energy gate. Processed frames $\tilde{x}_m[n]$ are \emph{overlapped and added} to form the output sequence $\tilde{x}[n]$.

\subsection{Normalization}
\label{sec:rms_normaliztion}

\gls{rms} normalization adjusts a signal’s dynamic range by scaling its energy to a predefined target level based on its \gls{rms} value. This technique is commonly used in speech detection and enhancement systems to ensure consistent signal energy levels, improving the reliability of energy-based features like short-time energy, which enhances speech detection robustness~\cite{rms_normalization_reference_1, rms_normalization_reference_2, rms_normalization_reference_3}.

To normalize a zero-mean signal $x[n]$ of length $N$, with a desired target RMS value $A$, the signal is first normalized by its computed \gls{rms} and then scaled to the target value, as expressed by
\begin{equation}
\label{formula:rms}
\text{\gls{rms}} = \sqrt{\frac{1}{N} \sum_{n=1}^{N}\!x^2[n]}~, \quad
\tilde{x}[n] = A \cdot \frac{x[n]}{\text{\gls{rms}}}.
\end{equation}
Moreover, to avoid clipping, the normalized signal is bounded within a valid amplitude range.

\section{Pipeline}
\label{sec:vad_pipeline}

The proposed \gls{vad} pipeline is constructed by integrating all the steps previously discussed. This framework processes an input audio signal through a sequential pipeline composed of segmentation, pre-processing, inference, grouping, and post-processing stages. An overview of the system is illustrated in Fig.~\ref{fig:vad_pipeline}.

The input audio signal is first divided into non-overlapping segments
\begin{equation}
    x = \left\{ x_{t} \right\}_{t=1}^{t=T},
\end{equation}
where $x_t$ denotes the $t$-th audio segment, and $T$ is the total number of segments. These segments are then processed by a \textit{pre-processing} module, which applies the pre-processing of Section~\ref{sec:preprocessing} to $x$ giving 
\begin{equation}
    x_{\text{pre}} = \left\{ x_{\text{pre}, t} \right\}_{t=1}^{t=T}.
\end{equation}

The sequence \( x_{\text{pre}} \) is passed to the \gls{sgvad} model that outputs a sequence of frame-wise predictions. Rather than relying on individual frame outputs, we utilize a sliding window of size $W$ to group consecutive predictions
\begin{equation}
    \hat{Y}_{\text{vad},t}=\left\{ \tilde{y}_{\text{sg}}(x_{\text{pre},k}) \right\}_{k=t}^{k=t+W-1}.
\end{equation}

The complete set of overlapping grouped predictions is denoted by
\begin{equation}
    \left\{ \hat{Y}_{\text{vad},t} \right\}_{t=1}^{t=T{-}W{+}1}.
\end{equation}

To obtain robust segment-level decisions, each group of predictions is processed by the \textit{post-processing} module, which applies the post-processing strategy detailed in Section~\ref{sec:postprocessing}. Specifically, it employs a majority voting mechanism to aggregate the predictions within each window \( \hat{Y}_{\text{vad},t} \). The majority voting is
\begin{equation}
    \begin{aligned}
        M_{t} 
              &= \mathbbm{1}_{\left\{\sum_{k=t}^{t+W-1} \!\tilde{y}_{\text{sg}}(x_{\text{pre},k}) > \left\lceil \frac{W}{2} \right\rceil \right\}} \\
              &~~~~~ \text{for} ~~~ t\!\in\!\{1, \dots, T\!-\!W\!+\!1\},
    \end{aligned}
\end{equation}
where $\lceil \cdot \rceil$ is the ceiling function. The output $M_t$ is set to 1 if more than half of the predictions in the window indicate speech activity; zero otherwise.

The overall \gls{vad} decision for the input audio is determined by checking whether any window has been classified as speech
\begin{equation}
    \tilde{y}_{\text{vad}} = \max_{t\in \{1,\ldots,T\}}{~M_{t}}=\!\begin{cases}
        1, & \begin{array}[t]{@{}r}
            \text{if}~~\exists~ t\in\{1, \dots, T{-}W{+}1\} \\
            \text{such that }M_t = 1,
            \end{array} \\
        0, & \begin{array}[t]{@{}l}
            \text{otherwise.}
        \end{array}
    \end{cases}
\end{equation}
This  binary label \( \tilde{y}_{\text{vad}} \in \{0, 1\} \) indicates the presence or absence of human voice activity in the entire audio signal.


\section{Experiments and Results}
\label{sec:discussion}
This section describes the datasets and experimental setup used to evaluate the proposed method.
\subsection{Datasets}
\label{sec:datasets}

We evaluate our \gls{vad} system using four diverse datasets, each containing varying speech and non-speech conditions, designed to test the model's performance under different scenarios.

\begin{itemize}
    \item \textbf{AVA Speech}: We use the AVA-Speech dataset~\cite{ava_speech_dataset} as it was used as an evaluation set in the \gls{sgvad} paper, which serves as our baseline. The dataset consists of audio clips from 160 YouTube videos, but due to some videos being deleted from the cloud, we were able to use a subset of the original dataset. Based on the provided annotations, we extracted 14,209 audio clips, with an average length of 4.5 seconds per clip, totaling about 17.9 hours of audio. 
    We categorize its data into three classes: \textit{clean speech}, \textit{noisy speech} (speech with noise or music), and \textit{non-speech}.

    \item \textbf{VBD Dataset (Voice Bank + DEMAND)}: The VBD dataset~\cite{VBD_dataset}, also known as Voice Bank + DEMAND, contains parallel recordings of clean and noisy speech at a sampling rate of 48 kHz, originally designed for training and testing speech enhancement methods. The dataset contains 19,748 audio clips, with an average length of 2.92 seconds per clip, totaling about 16.0 hours of audio. 
    
    \item \textbf{ESC50 Dataset (Environmental Sound Classification)}: The ESC50 dataset~\cite{ESC_50_dataset} contains 50 different environmental sound categories, such as animal calls, vehicle noises, and nature sounds. It is commonly used for environmental sound classification tasks. Notably, 10 of the 50 classes include human non-speech vocalizations, such as laughing, coughing, sneezing, and breathing, which are important to distinguish from actual speech. The dataset consists of 2,400 audio clips, with an average length of 5.0 seconds per clip, totaling 3.33 hours of audio. 

    \item \textbf{MS-SNSD Dataset (Microsoft Scalable Noisy Speech Detection)}: The MS-SNSD dataset~\cite{MS-SNSD_Microsoft_Noisy_dataset_framework} offers a comprehensive framework for generating noisy speech by combining clean speech recordings with various environmental noise types at \gls{snr} levels from 0 dB to 20 dB. This allows for the creation of a scalable noisy speech dataset tailored to specific requirements. By using this framework, we generated a dataset that includes 6,508 audio clips, with an average length of 2.90 seconds per clip, totaling about 5.2 hours of audio. 
    
\end{itemize}

All audio samples across these datasets are resampled to a unified sampling rate of 16\,kHz and segmented before being processed by our \gls{vad} pipeline.

\subsection{Experimental Settings}
\label{sec:settings}

The \gls{vad} system utilizes the inference model of \gls{sgvad} without modification, as depicted in Fig.~\ref{fig:vad_pipeline}. Each audio input is segmented into 200 ms chunks, on which the model performs inference to detect speech activity. For post-processing, we apply the majority voting rule 3 out of 4, meaning that a window is classified as speech if at least three out of the four chunks are predicted as speech. We evaluated our proposed method under three settings: (1) \gls{sgvad} as the baseline, (2) \gls{sgvad} with post-processing (majority voting), referred to as \gls{vad}~1, and (3) \gls{sgvad} with both post-processing and pre-processing (noise removal), referred to as \gls{vad}~2.

The performance is evaluated using three key metrics. The first metric is the \gls{roc} curve, which assesses the trade-off between \gls{tpr} and \gls{fpr} at varying thresholds, offering a comprehensive evaluation of model performance. The second metric is the \gls{auc}, which summarizes the overall performance of the model by calculating the area under the \gls{roc} curve. The third metric is detection accuracy, computed for each of the three classes: \textit{clean speech}, \textit{noisy speech}, and \textit{non-speech}, by comparing predicted labels with ground truth annotations of each class.

\subsection{Results}
\label{sec:results}

We report the performance of three \gls{vad} models, as detailed in Section~\ref{sec:settings}, across four distinct datasets described in Section~\ref{sec:datasets}. The detection accuracy for each model, evaluated on the categories of clean speech, noisy speech, and non-speech, is presented in Table~\ref{tab:detection_accuracies}.

The accuracy results for noisy speech highlight a notable improvement in favour of our models across different datasets. Bold values in the table indicate the highest accuracy achieved for a given dataset. Although the baseline model (\gls{sgvad}) consistently performs well in detecting non-speech (noise) segments, it exhibits lower accuracy in detecting noisy speech and clean speech on certain datasets, particularly those that lack continuous human speech (e.g., VBD and MS-SNSD).

Specifically, the \gls{sgvad} baseline model struggles with detecting noisy speech across all datasets and encounters challenges with clean speech detection in datasets characterized by more complex sound environments, such as MS-SNSD. In contrast, \gls{vad}~1 that combines \gls{sgvad} with post-processing demonstrates significant improvements in both noisy speech and clean speech detection, while maintaining comparable performance in non-speech detection to the baseline. This clarifies the effectiveness of the majority voting rule in enhancing performance for speech detection. Moreover, \gls{vad}~2, which includes both pre-processing and post-processing with \gls{sgvad} shows substantial improvements in detecting both clean and noisy speech.
\begin{table}[t]
\centering
\caption{Detection accuracy in percentages, the higher the better, of evaluated \gls{vad} models across four datasets and three audio categories: clean speech, noisy speech, and non-speech. Bold values indicate the best performance per row. \gls{vad}~2 shows consistent improvements in speech detection, particularly under noisy conditions.}
\resizebox{0.95\columnwidth}{!}{
{\small
\begin{tabular}{llccc}
\toprule
\textbf{Dataset} & \textbf{Type} &
\makecell{\textbf{\gls{sgvad}} \\ Baseline} &
\makecell{\textbf{\gls{vad}~1} \\ Majority \\ (3 out of 4)} &
\makecell{\textbf{\gls{vad}~2} \\ Majority \\ (3 out of 4) \\ + Preproc.} \\
\midrule

\multirow{3}{*}{AVA~\cite{ava_speech_dataset}} 
  & Non-Speech    & \textbf{99.7\%} & 97.4\% & 81.3\% \\
  & Clean Speech  & 74.7\% & 79.9\% & \textbf{92.1\%} \\
  & {\bf Noisy Speech}  & 45.4\% & 64.9\% & \textbf{84.7\%} \\

\midrule

\multirow{2}{*}{VBD~\cite{VBD_dataset}} 
  & Clean Speech & 66.8\% & 98.3\% & \textbf{99.1\%} \\
  & {\bf Noisy Speech} & 32.5\% & 73.7\% & \textbf{97.4\%} \\

\midrule

ESC50~\cite{ESC_50_dataset} & Non-Speech & \textbf{98.9\%} & 93.2\% & 85.5\% \\

\midrule

\multirow{3}{*}{\makecell{MS-SNSD~\cite{MS-SNSD_Microsoft_Noisy_dataset_framework}}} 
  & Non-Speech    & \textbf{99.1\%} & 97.9\% & 87.7\% \\
  & Clean Speech  & 58.7\% & 97.4\% & \textbf{98.7\%} \\
  & {\bf Noisy Speech}  & 15.9\% & 66.4\% & \textbf{89.9\%} \\

\bottomrule
\label{tab:detection_accuracies}
\end{tabular}
}
}
\end{table}

\begin{figure}[t!]
    \centering
    \includegraphics[width=0.84\linewidth]{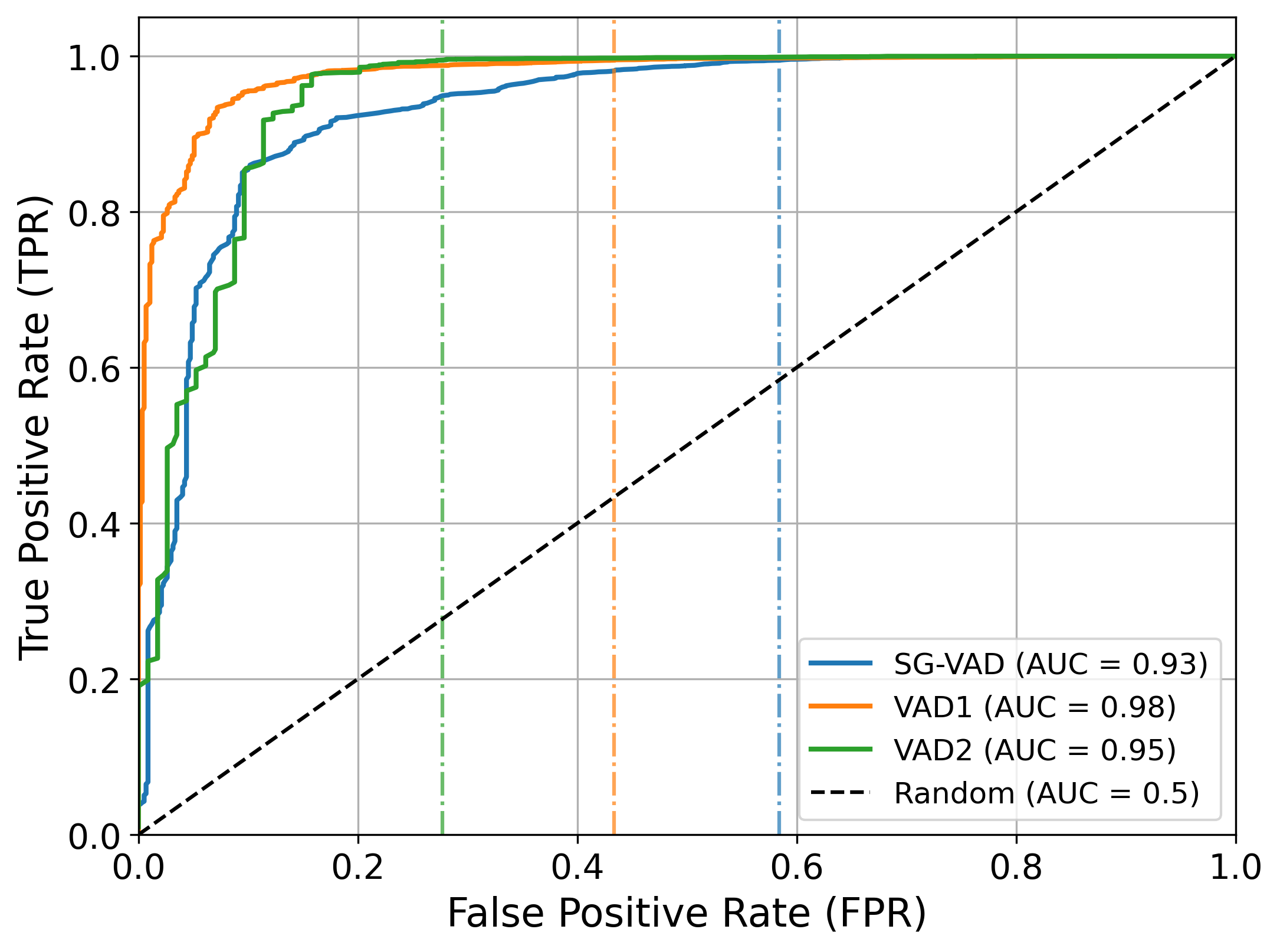}
    \caption{\Gls{roc} curves for \gls{sgvad} (baseline), \gls{vad}~1 (with post-processing), and \gls{vad}~2 (with post-processing and pre-processing). \Gls{vad}~2 achieves the lowest \gls{fpr} (28\%) at 99\% \gls{tpr}, compared to \gls{vad}~1 (43\%) and the baseline (58\%), highlighting its advantage in reducing false positives.
    }
    \label{fig:roc_curves}
\end{figure}

The \gls{roc} curve in Fig.~\ref{fig:roc_curves} corresponds to a binary classification experiment where all speech segments (both clean and noisy) are grouped into a single \textit{speech} class and evaluated against the \textit{non-speech} (noise) class. This figure reveals that the \gls{auc} for \gls{vad}~1 (\gls{auc} = 0.98) and \gls{vad}~2 (\gls{auc} = 0.95) both exceed the baseline model (\gls{auc} = 0.93). While \gls{auc} is a valuable metric, the most critical aspect in a voice assistant pipeline is minimizing the \gls{fpr} while maintaining a high \gls{tpr}. False positive reduction is particularly important, as the goal of a \gls{vad} system in such applications is to ensure that nearly all speech activity is detected (high \gls{tpr}) while minimizing instances where non-speech segments are mistakenly classified as speech (low \gls{fpr}).

As illustrated in the \gls{roc} curve, \gls{vad}~2 achieves a \gls{tpr} of 99\% at an \gls{fpr} of 28\%, while the baseline model achieves the same \gls{tpr} but with a much higher \gls{fpr} of 58\%. \gls{vad}~1 outperforms the baseline, with a \gls{tpr} of 99\% at an \gls{fpr} of 43\%, but still lags behind \gls{vad}~2. Overall, for a  high \gls{tpr} at the level of $99\%$, \gls{vad}~2 offers a significantly lower \gls{fpr}, making it more suitable for voice activity detection in voice assistants, where reducing false positives is the key objective.

\section{Conclusion}
\label{sec:conclusion}
Commonly-used voice assistants struggle with detection of human voice in the presence of background noise. Handling complex noise in the background often requires  large models, that limits their deployment on low-resource edge devices.  By adding data pre- and post-processing modules we transform a light-weight \gls{vad} trained on clean speech to a noise-robust version. These pre- and post- processing steps also improve the accuracy of clean human voice detection. 



\begin{thebibliography}{10}

\bibitem{benyassine1997itu}
A.~Benyassine et~al.,
\newblock ``{ITU-T} recommendation g. 729 annex {B}: a silence compression scheme for use with g. 729 optimized for v. 70 digital simultaneous voice and data applications,''
\newblock {\em IEEE Communications Magazine}, vol. 35, no. 9, pp. 64--73, 1997.

\bibitem{sehgal2018convolutional}
Abhishek Sehgal and Nasser Kehtarnavaz,
\newblock ``A convolutional neural network smartphone app for real-time voice activity detection,''
\newblock {\em IEEE Access}, vol. 6, pp. 9017--9026, 2018.

\bibitem{chang2018temporal}
S.~Y.~Chang et~al.,
\newblock ``Temporal modeling using dilated convolution and gating for voice-activity-detection,''
\newblock in {\em International Conference on Acoustics, Speech and Signal Processing}. IEEE, 2018, pp. 5549--5553.

\bibitem{braun2021training}
Sebastian Braun and Ivan Tashev,
\newblock ``On training targets for noise-robust voice activity detection,''
\newblock in {\em European Signal Processing Conference}. IEEE, 2021, pp. 421--425.

\bibitem{jia2021marblenet}
Fei Jia, Somshubra Majumdar, and Boris Ginsburg,
\newblock ``Marblenet: Deep 1d time-channel separable convolutional neural network for voice activity detection,''
\newblock in {\em International Conference on Acoustics, Speech and Signal Processing}. IEEE, 2021, pp. 6818--6822.

\bibitem{SG_VAD_main_paper}
Jonathan Svirsky and Ofir Lindenbaum,
\newblock ``{SG-VAD}: Stochastic gates based speech activity detection,''
\newblock in {\em International Conference on Acoustics, Speech and Signal Processing}, 2023, pp. 1--5.

\bibitem{gscv2_dataset}
Pete Warden,
\newblock ``Speech commands: A dataset for limited-vocabulary speech recognition,'' 2018.

\bibitem{fs2k_dataset}
Frederic Font, Gerard Roma, and Xavier Serra,
\newblock ``Freesound technical demo,''
\newblock in {\em Proceedings of the 21st ACM International Conference on Multimedia}, New York, NY, USA, 2013, MM '13, p. 411–412, Association for Computing Machinery.

\bibitem{ava_speech_dataset}
S.~Chaudhuri et~al.,
\newblock ``Ava-speech: A densely labeled dataset of speech activity in movies,''
\newblock in {\em Proceedings of Interspeech}, 2018.

\bibitem{havic_dataset}
S.~Strassel et~al.,
\newblock ``Creating havic: Heterogeneous audio visual internet collection,''
\newblock in {\em Proceedings of the Eighth International Conference on Language Resources and Evaluation}, 2012, pp. 2573--2577.

\bibitem{non-continuous_speaking_supporting_reference}
F.~L. I.~Dutsinma et~al.,
\newblock ``A systematic review of voice assistant usability: An iso 9241–11 approach,''
\newblock {\em Springer Nature Computer Science Journal}, vol. 3, no. 4, 2022.

\bibitem{spectral_subtraction_reference_1}
S.~Boll,
\newblock ``Suppression of acoustic noise in speech using spectral subtraction,''
\newblock {\em IEEE Transactions on Acoustics, Speech, and Signal Processing}, vol. 27, no. 2, pp. 113--120, 1979.

\bibitem{spectral_subtraction_reference_2}
M.~Berouti, R.~Schwartz, and J.~Makhoul,
\newblock ``Enhancement of speech corrupted by acoustic noise,''
\newblock in {\em International Conference on Acoustics, Speech, and Signal Processing}, 1979, vol.~4, pp. 208--211.

\bibitem{spectral_subtraction_reference_3}
Philipos~C. Loizou,
\newblock {\em Speech enhancement: theory and practice},
\newblock CRC press, 2013.

\bibitem{energy_gating_reference_1}
L.~R. Rabiner and M.~R. Sambur,
\newblock ``An algorithm for determining the endpoints of isolated utterances,''
\newblock {\em The Bell System Technical Journal}, vol. 54, no. 2, pp. 297--315, 1975.

\bibitem{energy_gating_reference_2}
K.~H.~Woo et~al.,
\newblock ``Robust voice activity detection algorithm for estimating noise spectrum,''
\newblock {\em Electronics Letters}, vol. 36, pp. 180 -- 181, 02 2000.

\bibitem{rms_normalization_reference_1}
Aaron~E. Rosenberg, Chin-Hui Lee, and Frank~K. Soong,
\newblock ``Cepstral channel normalization techniques for hmm-based speaker verification,''
\newblock in {\em International Conference on Spoken Language Processing}, 1994.

\bibitem{rms_normalization_reference_2}
Philipos~C. Loizou,
\newblock {\em Speech Enhancement: Theory and Practice},
\newblock CRC Press, 2nd edition, 2013.

\bibitem{rms_normalization_reference_3}
Lawrence~R. Rabiner and Ronald~W. Schafer,
\newblock {\em Introduction to Digital Speech Processing}, vol.~1,
\newblock Foundations and Trends in Signal Processing, 2007.

\bibitem{VBD_dataset}
C.~Valentini~Botinhao et~al.,
\newblock ``Speech enhancement for a noise-robust text-to-speech synthesis system using deep recurrent neural networks,''
\newblock in {\em Proceedings of Interspeech}, Sept. 2016, pp. 352--356.

\bibitem{ESC_50_dataset}
Karol~J. Piczak,
\newblock ``{ESC}: {Dataset} for {Environmental Sound Classification},''
\newblock in {\em Proceedings of the 23rd {Annual ACM Conference} on {Multimedia}}. 2015, pp. 1015--1018, {ACM Press}.

\bibitem{MS-SNSD_Microsoft_Noisy_dataset_framework}
C.~KA~Reddy et~al.,
\newblock ``A scalable noisy speech dataset and online subjective test framework,''
\newblock {\em Proceedings of Interspeech}, pp. 1816--1820, 2019.

\end{thebibliography}

\end{document}